\documentclass[aps,prl,preprint,groupedaddress,longbibliography]{revtex4-1}

\usepackage{color}
\usepackage{graphicx}
\usepackage{amsmath}
\usepackage{bm}
\usepackage{amsfonts}
\usepackage{amssymb}

\newcommand{\ma}{\textrm{\textbf{\textit{a}}}}
\newcommand{\mb}{\textrm{\textbf{\textit{b}}}}
\newcommand{\mc}{\textrm{\textbf{\textit{c}}}}

\begin{document}
\raggedbottom


\title{The reconfigurable Josephson circulator/directional amplifier}


\author {K.~M. Sliwa$^{\ast}$ and M. Hatridge, A.~Narla, S.~Shankar,  L. Frunzio, R. J. Schoelkopf, M.~H. Devoret$^{\ast}$\\
\normalsize{Departments of Applied Physics and Physics, Yale University, New Haven, Connecticut}\\
\normalsize{$^\ast$To whom correspondence should be addressed: E-mail:  katrina.sliwa@yale.edu, michel.devoret@yale.edu}
}
\affiliation{}


\date{\today}
\begin{abstract}
Circulators and directional amplifiers are crucial non-reciprocal signal routing and processing components involved in microwave readout chains for a variety of applications. They are particularly important in the field of superconducting quantum information, where the devices also need to have minimal photon losses to preserve the quantum coherence of signals. Conventional commercial implementations of each device suffer from losses and are built from very different physical principles, which has led to separate strategies for the construction of their quantum-limited versions. However, as recently proposed theoretically, by establishing simultaneous pairwise conversion and/or gain processes between three modes of a Josephson-junction based superconducting microwave circuit, it is possible to endow the circuit with the functions of either a phase-preserving directional amplifier or a circulator. Here, we experimentally demonstrate these two modes of operation of the same circuit. Furthermore, in the directional amplifier mode, we show that the noise performance is comparable to standard non-directional superconducting amplifiers, while in the circulator mode, we show that the sense of circulation is fully reversible. Our device is far simpler in both modes of operation than previous proposals and implementations, requiring only three microwave pumps. It offers the advantage of flexibility, as it can dynamically switch between modes of operation as its pump conditions are changed. Moreover, by demonstrating that a single three-wave process yields  non-reciprocal devices with reconfigurable functions, our work breaks the ground for  the development of future, more-complex directional circuits, and has excellent prospects for on-chip integration.

\end{abstract}

\maketitle


%

\section {Introduction}

Devices that violate reciprocity, the symmetry that exists upon exchange of source and observer, are of great practical and theoretical importance. They allow, for example, the control of information flow in signal processing applications \cite{MPozar2005} and violations of detailed balance\cite{Snyder98, Zhu2014}. In the microwave domain, two ubiquitous and vital directional elements are the matched directional amplifier  and the circulator \cite{MPozar2005}. These elements generally function by very different underlying physical phenomena. Amplifiers are typically made with transistors which use the field effect of a metallic gate on a semiconducting channel to unidirectionally increase the power of a weak input signal (Fig.~1A).  On the other hand, microwave circulators usually operate by the interference of multiple paths through a magnetically biased ferrite ring which produces non-reciprocal phase shifts (Fig.~1B). Both these elements are widely used for the processing of microwave signals in superconducting qubit experiments \cite{Clarke2008, Devoret2013}. However, commercially available devices are not ideal for quantum information applications since they suffer from losses, and associated added noise, which harms the quantum properties of flying microwave photons \cite{Hatridge2013,Murch2013,Roch2014}. Additionally, the required strong magnetic fields prevent easy on-chip integration of circulators with superconducting qubits. The rather distinct physical phenomena underlying current commercial implementations of circulators and amplifiers would suggest that these directional devices must be improved by completely different means.

Here, we demonstrate that both kinds of non-reciprocal functions can be produced in the same device in a dispersive manner via the interference of parametric couplings  between multiple resonant structures (Fig.~1C). Previously, several circulator designs without magnetic elements have been proposed \cite{Kamal2010, Lehur2010, Kerckhoff2015}, tailored for superconducting qubit experiments. An RF-frequency circulator with only semiconducting parts has also been recently demonstrated experimentally \cite{Estep2014}, but using lossy components. The proposed designs specific to superconducting qubit  experiments are all based on parametrically pumping a distributed nonlinearity with several RF-sources to produce unity-gain photon conversion between the ports of the device. These parametric couplings can in fact produce photon number gain as well. The interference of this gain process with unity-gain photon conversion has been discussed in several recent theory proposals \cite{Metelmann2014, Ranzani2015, Metelmann2015} as a means to realize directional amplification. In particular, the authors of Ref.~\onlinecite{Ranzani2015} have developed a remarkable graph-theoretical scheme to predict the minimal non-reciprocal circuit satisfying a given set of constraints. In this work we experimentally realize a directional amplifier based on this principle, suitable for quantum information purposes. We further show that via a simple change in pump conditions, we can switch the device into a purely dispersive, ferrite-free microwave circulator which is truly minimal, unlike previous proposals. In the process we demonstrate the deep physical connection between these two seemingly distinct directional elements.

\section{Theory of the circulator and directional amplifier based on the Josephson Parametric Converter (JPC)}

Our device is realized by parametrically pumping the three wave mixing non-linearity of a Josephson Parametric Converter (JPC) \cite{Bergeal2010, Bergeal2010a}. The heart of the JPC is the Josephson Ring Modulator (JRM), a ring of four Josephson junctions, which couples to three orthogonal circuit modes $\ma$, $\mb$, and $\mc$ shown schematically in Fig.~2A. The JRM is embedded at the central current antinodes of two crossed $\lambda/2$-resonators, which constitute modes \ma~and \mb, with frequencies $\omega_{\ma}/2\pi=9.167$~GHz, $\omega_{\mb}/2\pi=5.241$~GHz. These resonators are accessible by ``ports'', i.e.\  semi-infinite transmission lines giving the modes finite energy decay rates $\kappa_\ma/2\pi=44$~MHz, $\kappa_\mb/2\pi=19$~MHz respectively. The third microwave mode \mc, with frequency $\omega_{c}/2\pi=7.174$~GHz, is formed by the common-mode excitation of the JRM, and has an energy decay rate $\kappa_\mc/2\pi=50$~MHz. On applying a flux through the JRM close to half a flux quantum, a tri-linear three-body interaction between these three orthogonal modes of the form $\mathcal{H}_{int}=g_{3}\left(\ma^\dag\mb^\dag\mc+\ma\mb\mc^\dag\right)$ becomes the leading non-linear term in the system hamiltonian\cite{Abdo2013a}. Here \ma, \mb, and \mc~are the annihilation operators for their respective modes, and $g_{3}$ is the flux-dependent tri-linear coupling strength. By off-resonantly driving one mode with a CW tone of the appropriate frequency, we can produce one of two kinds of two-body interactions between the other two modes as discussed below.

The first kind of 2-body interaction is a photon gain process, yielding the  conventional non-directional phase-preserving amplification which has been widely used previously \cite{Bergeal2010a,  Flurin2012, Hatridge2013, Campagne-Ibarcq2013}. This process is achieved by pumping one spatial mode (e.g.\ \mb) at the sum frequency of the other two ($\omega_{\mb}^p=\omega_{\ma}+\omega_{\mc}$).  Provided the pump frequency is sufficiently detuned from any harmonic of the \mc-mode, the pump can be approximated as a classical drive and the interaction Hamiltonian reduces to $\mathcal{H}_{int}^G=|g_{\ma \mc}| (e^{+i \phi_p}\ma^\dag \mc^\dag +e^{-i \phi_p}\ma \mc )$, where $g_{\ma \mc}$ is the pump-power-dependent coupling strength, and $\phi_p$ is the pump phase, referred to a common clock used for all CW tones. The gain of the resulting amplification process at zero detuning can be written as $\sqrt{G}=(1+|g_{\ma \mc}|^2/\kappa_{\ma}\kappa_{\mc})/(1-|g_{\ma \mc}|^2/\kappa_{\ma}\kappa_{\mc})$. Classically, as $|g_{\ma \mc}|^2 \to\kappa_\ma \kappa_\mc$, the small signal gain can become arbitrarily large.  This process, denoted graphically as `$G$', bi-directionally connects two modes, up to a phase factor, as shown schematically in Fig.~2B. We also give a graphical representation of the scattering matrix in Fig.~2B.  Signals entering through one port are amplified in reflection with voltage gain $\sqrt{G}$, and in transmission with gain $\sqrt{G-1}$, together with frequency translation and a pump-phase dependent non-reciprocal phase shift $\phi_p$. Phase conjugation is also taking place in the frequency conversion process, as indicated by a white (rather than black) arrowhead.  A signal incident on one mode will be combined with amplified vacuum-fluctuations from the other mode, achieving phase-preserving amplification.  Due to the symmetry of the amplification process, the signal can be collected from either output port.

The second form of 2-body interaction we employ is a unity-gain photon conversion process \cite{Abdo2013}.To demonstrate that any pair of modes can support either two-body interaction, here we chose to couple modes $\ma$ and $\mb$. This is achieved by pumping the third mode $\mc$ at the difference frequency of the other two ($\omega_{\mc}^p=\omega_{\ma}-\omega_{\mb}$). Again, provided the pump frequency is sufficiently detuned from any mode, the interaction Hamiltonian reduces to $\mathcal{H}_{int}^C=|g_{\ma \mb} |(e^{+i \phi_p}\ma \mb^\dag +e^{-i \phi_p}\ma^\dag \mb )$. The process is characterized by the conversion coefficient at zero detuning $C=(4 |g_{\ma \mb}|^2/\kappa_\ma \kappa_\mb)/(1+ |g_{\ma \mb}|^2/\kappa_\ma \kappa_\mb)^2$, ranging from 0 (no-conversion) to 1 (full-conversion when $|g_{\ma \mb}|=\sqrt{\kappa_\ma \kappa_\mb}$)\cite{Abdo2013}.  The process schematic is shown in Fig.~2C, together with a graphical representation of the scattering matrix.  Signals incident on either mode are either reflected with coefficient $\sqrt{1-C}$), or transmitted with coefficient $\sqrt{C}$. As in the gain process, signals transmitted through the device experience frequency translation and a pump-phase dependent non-reciprocal shift (note that here all arrowheads are black, as no phase conjugation occurs).  At the full conversion working point, $C=1$, the device resembles a gyrator\cite{MPozar2005} but with the complication that it also performs a frequency translation.


As shown in Fig.~2D, the three modes of the JPC can be connected with up to six simultaneous gain and conversion processes.  A particular function can be realized by identifying the appropriate pumping configuration.  Here, we will focus on configurations based only on one process per pair of modes, calculating the scattering matrices following the method of Ref. \onlinecite{Ranzani2015}. We first consider the case of three simultaneous unity-gain photon conversion processes, which produces a circulator.  A schematic for the device coupling configuration and a graphical representation of the scattering matrix are shown in Fig.~3A. In the ideal case, the circulator uses three conversion processes, which would each individually achieve full conversion.  The final important control variable is the algebraic sum of the three pump phases ($\phi_{tot}^{circ} = \phi_\ma +\phi_\mb - \phi_\mc$). Constructive or destructive interference occurs as signals travel around the device, their phase being controlled by $\phi_{tot}^{circ}$, which acts here as an artificial gauge flux, and plays the role of the magnetic field in a conventional circulator. As shown in Fig.~3B, for $\phi_{tot}^{circ}=\pm \pi/2$, a matched circulator with clockwise/counter-clockwise circulation is created.  We note that this circulator is somewhat different from a ferrite-based circulator in that it translates the frequencies of signals passing through it, but this is generally not detrimental as we can freely shift the carrier frequency of our microwave signals without degrading their information content. As a matter of fact, frequency translation is beneficial in amplification chains when unwanted feedback needs to be suppressed. 

Directional amplification is achieved by combining two gain processes and one conversion process, as shown schematically in Fig.~4A. In the ideal case, we set the pairwise processes so that we have two equal gains $G$ and one full conversion ($C=1$).  Again, the interference within the device is controlled by a total pump phase $\phi_{tot}^{d\text{-}a}$, here now given by $\phi_\ma-\phi_\mb+\phi_\mc$, with directional amplification occurring at $\phi_{tot}^{d\text{-}a} = \pm \pi /2$. Unlike the circulator, this schematic has a pronounced asymmetry in signal flow through the device, as shown by the graphical scattering matrix in Fig.~4B.  We therefore label the three ports in the directional amplifier by the roles played by their inputs as the Signal (S) input, Idler (I) input, and Vacuum (V) input. Signals incident on S correspond to the directional amplifier input, while signals exiting I form its output. The S port is matched (no power reflects), and incident power is instead transmitted with gain to the I and V ports. Vacuum fluctuations incident on I are responsible for the additional half-photon of added quantum noise necessarily associated with quantum-limited phase-preserving amplification. Signals incident on the V port are noiselessly and directionally transmitted through the device to the S port with unity photon gain. Any device must necessarily reflect, at a minimum, vacuum fluctuations back to the upstream signal source. In our implementation the source of these fluctuations would be provided by the cold $50$~$\Omega$ load terminating the V port. Furthermore, the unity-gain transmission of signals from the V to the S port follows from the combined requirements of quantum-limited amplification (sending it to I would degrade the noise performance of the device) and the information conserving nature of the device (no entropy produced since we assume the pump to be perfectly stiff)\cite{Abdo2013a}.  Changing $\phi_{tot}^{d\text{-}a}$ by $\pi$ flips which physical port plays the role of S and V, with I remaining unchanged.  The roles can be further re-mapped by changing which pair of modes is linked via conversion and thus, in general, each of the three physical ports can play each role.  We note that combining two gain processes with gain $G$ yields a directional amplifier with gain $G$, not $G^2$.  The combined operation should be thought of as rerouting the outputs of one port of a non-directional amplifier (from S to V) rather than as two independent stages of amplification.


Unlike the circulator, which is relatively forgiving of imperfect conversion ($C < 1$), the directional amplifier is much less tolerant of non-idealities. Its directionality is only achieved when the conversion process is adjusted so that $1-C <1/G$ (see supplementary material). This behavior is demonstrated theoretically in Fig.~5A. The gains have been set to a finite value of $\sim12$~dB, and the input match, initially perfect, degrades as the conversion coefficient decreases. For any pair of finite gains, there exists a conversion coefficient ($C=0.95$ for $G=12$~dB) below which any semblance of directional amplification is lost.



\section{Experimental apparatus and results}

Fig.~2A shows a schematic of our experimental setup anchored to the mixing chamber of a dilution refrigerator operated below $30$~mK.  A detailed diagram of the experimental setup, including both the details of the measurement setup with a 2-port vector network analyzer (VNA) and the fridge wiring, can be found in Supp. Fig.~S4. In the experiment, our JPC is slightly modified from the generic one, having an $8$-junction JRM.  The inner four junctions do not participate in the coupling Hamiltonian, instead increasing the device's tuneability and providing stability near the half-flux-quantum working point \cite{Roch2012}.  The frequencies of all three modes can be tuned over a 400 MHz span by varying the external magnetic flux through the ring. For this experiment we have chosen a flux such that mode $\ma$ is at $\omega_{\ma}/2\pi=9.167$~GHz with $\kappa_\ma/2\pi=44$~MHz , mode \mb\ is at $\omega_{\mb}/2\pi=5.241$~GHz with $\kappa_\mb/2\pi=19$~MHz, and mode \mc\ is at $\omega_{c}/2\pi=7.174$~GHz with $\kappa_\mc/2\pi=50$~MHz. The JPC is connected via cascaded $180^{\circ}$ hybrids (Krytar 4040124) in order to separately address all three modes. Pumps and probe tones are applied to each mode via the weakly coupled port of a directional coupler (Krytar 104020020) connected to each hybrid. For probing the device in the circulator mode, the signal power at the output of the VNA of $-55$~dBm was applied, right before saturation effects appeared (see Supp. Fig.~S4 for line attenuations). For probing the device in the directional amplifier mode, the signal power at the output of the VNA was $-75$~dBm, which was well out of the saturated regime. The difference in probe power for these two modes is expected since the output power of the device must remain the same in both cases. We verified that in the absence of any pump the modes are completely orthogonal, for example, when probing mode $\ma$, no response is seen at the frequency of modes $\mb$ and $\mc$. All pumps were generated by physically separate generators locked to a common $10$~MHz rubidium frequency standard, ensuring the stability of the total pump phase during an experimental run.

\subsection{Circulator}

We realize a Josephson circulator by coupling all three modes pairwise via conversion processes as previously described. When the total pump phase is set to $\phi_{tot}^{circ}=\pi/2$, circulation is clockwise from mode \ma\ to \mb\ to \mc\, as shown in Fig.~3A.  When $\phi_{tot}^{circ}$ is changed by $\pi$, with no other changes to pump parameters, the sense- of circulation is reversed (Fig.~3A). The pump frequencies and powers at their respective generator outputs were $\omega^p_{\mc}/2\pi=3.928$~GHz, $P^p_{\mc}=-29.92$~dBm, $\omega^p_{\ma}/2\pi=1.9291$~GHz, $P^p_{\ma}=-7.42$~dBm, and $\omega^p_{\mb}/2\pi=1.9989$~GHz, $P^p_{\mb}=1.9$~dBm, corresponding to conversion coefficients $C$ of $0.97$, $0.98$, and $0.99$, respectively. These processes were first applied singly, and the resultant conversion processes were compared to collectively maximize the conversion coefficients while simultaneously matching the single mode frequency responses of all three modes (Supp. Fig.~S1). The magnitude of the conversion coefficients were then fine-tuned by maximizing the magnitude and symmetry of both the input match and the reverse isolation of all three ports of the circulator.



Figure 3B shows the complete set of measured scattering parameters ($s_{ij}$, $i,j = \ma,\mb,\mc$) for the circulator as a function of probe frequency. For the diagonal components of the scattering matrix, the reflected output is directed to an amplifier chain and recorded at room temperature by the VNA. Off-diagonal components of the scattering matrix involve inputs and outputs which are at different frequencies, and therefore the output tones are mixed with a local oscillator at the corresponding difference frequency to translate them back to the probe tone frequency prior to being acquired by the VNA. 

We identify the unknown offset in the total phase $\phi_{tot}^{circ}$ by finding the two values, differing by $\pi$, for which $s_{\mb\mb}$ is minimized. Following the convention in the theory section, we assign $\phi_{tot}^{circ}=\pi/2$ to clockwise circulation.  We note that it suffices to vary only one pump phase to achieve a desired $\phi_{tot}^{circ}$, and so only the phase of the pump applied to \mc\ was varied for all data shown in this paper, although we verified in a separate experiment that the response to all three pump phases was equivalent.

As seen in Fig.~3B, on resonance we have a matched device (with reflection better than $-10$~dB) exhibiting more than $18.5$~dB reverse isolation, and less than $0.5$~dB of insertion loss.  The insertion loss is calibrated relative to the three individual conversion processes, which have been previously demonstrated to be efficient to within $0.1$~dB\cite{Abdo2013}.  Off resonance, the bandwidth of the individual conversion processes which comprise the circulator combine to give an $11$~MHz bandwidth over which the input match of all ports is better than $-10$~dB and the insertion loss is better than $1$~dB.

Simply flipping the pump phase by $\pi$ to $\phi_{tot}^{circ}=-\pi/2$, without any other variation of pump parameters, switches the direction of circulation, as shown. We see no degradation in overall device performance, and good agreement with theoretical calculations for the scattering parameters in both directions, given that the only input the theory uses is the three mode bandwidths and the conversion coefficients of the three individually pumped conversion processes.  We note that most deviations are associated with signals input to mode \mc.  We attribute these to the degradation in the spatial mode matching due to phase mismatches in the three cascaded hybrids versus \ma\ and \mb\ which each pass through a single hybrid.  The overall device performance is limited by imperfections in the pairwise conversion processes and drift in the overall pump phase.

We also characterized the device by measuring two representative scattering parameters, $s_{\mb\mb}$ and $s_{\mc \mb}$, as a continuous function of pump phase (Fig.~3C). The data are in excellent agreement with theory, showing three working points with alternating circulation directions at points separated by $\pi$ in phase ($-3\pi/2,-\pi/2, \pi/2$), with smooth transitions in the scattering parameters versus frequencies in between. Further experimental and theoretical work are required to predict and characterize the effect of higher order nonlinearities on the fine details of the device performance. This is especially vital for determining how many probe photons the device can process without degradation of performance.

\subsection{Directional amplifier}
As mentioned earlier, we realize a directional amplifier by setting, in the triangle of modes, two legs to conjugating photon gain processes and the remaining leg to a unity-gain photon conversion process (Fig.~4A). Modes \ma\ and \mb\ are coupled so that $C = 0.998$ via a pump applied at $\omega^p_{\mc}/2\pi=3.927$~GHz with a power of $P^p_{\mc}=-28.95$~dBm at the generator output, modes \ma\ and \mc\ with $G=13$ dB via a pump at $\omega^p_{\mb}/2\pi=16.339$~GHz with $P^p_{\mb}=-11.77$~dBm, and modes \mb\ and \mc\ with $G=12$ dB via a pump at $\omega^p_{\ma}/2\pi=12.412$~GHz with $P^p_{\ma}=-18.53$~dBm. These values were chosen experimentally both to approach perfect conversion and to minimize frequency offsets in the single pump mode responses subject to the constraint $\omega^p_{\mc}=\omega^p_{\mb}-\omega^p_{\ma}$ (Supp. Fig.~S2). Here, as with the circulator we remove offsets in the pump phases by finding $\phi_{tot}^{d\text{-}a}$ values for which $s_{\mc\mc}$ is minimized.  We again  define $\phi_{tot}^{d\text{-}a}$ to be $\pm \pi/2$ at these points, the sign being set by the direction of amplification.

In this mode of operation, physical ports can take on different roles depending on which modes are coupled via conversion and on the value of $\phi_{tot}^{d\text{-}a}$. Given our pump frequency configuration, when we set  $\phi_{tot}^{d\text{-}a}=-\pi/2$, mode \ma\ is the signal port S, mode \mb\ the vacuum port V, and mode \mc\ the idler port I as shown in the Fig.~4A. The scattering parameters are plotted in Fig.~4B, showing all the hallmarks of directional amplification. First, the input ports S and V show a reflection coefficient of $-16$~dB or greater, indicating the device is matched, while the third port shows gain in reflection.  Next, signals input at S are amplified and transmitted to I and V (gain of $14$~dB).  Third, signals incident on I are isolated from S (with isolation of $8$~dB), and are instead reflected from I and transmitted to V with gain.  Finally, signals incident on V are transmitted with near unity photon gain to S ($s_{\ma\mb}=0.2$~dB).  In normal operation, port V will be terminated in a cold $50$~$\Omega$ load and can be seen as providing the necessary vacuum fluctuations which must be emitted from S. The directional gain falls off with probe frequency as a lorentzian lineshape with a $3$-dB bandwidth of $11$~MHz, though we note that other bandwidths can be defined based on the required input match or reverse isolation.

Changing the total pump phase to $\phi_{tot}^{d\text{-}a}=\pi/2$ switches the roles of mode \ma\ and mode \mb. This is most directly seen by comparing $s_{\ma\mb}$ and $s_{\mb\ma}$, in which the direction of the gain reverses. As shown by the correspondence with the theory curves, all the other scattering parameters also change as expected. Again, the theory curves are calculated with the 3 mode bandwidths and the individual gain and conversion coefficients. In general, the agreement is not as good as for the circulator which we attribute to the fact that there is now gain in the system and therefore misalignments of the pairwise processes  and phase drifts can more drastically affect the amplifier performance. In practical implementations, interferometric techniques could be used to stabilize $\phi_{tot}^{d\text{-}a}$, or several matched, low-gain stages could be cascaded to achieve high net gain without  requiring extreme pump precision. As with the circulator, more sophisticated theoretical analysis is needed to understand the effects of higher-order nonlinearities. The effect of such nonlinearities can be seen, for instance, in distortions of the conversion process shown in Supp.~Fig.~S2. These are also crucial to understanding the dynamic range of the amplifier.

One of the most important characteristics of such paramps is the noise performance. This is characterized by the Noise Visibility Ratio (NVR) of the device, defined as the excess noise visible in a spectrum analyzer at room temperature when the amplifier is turned on versus off. This technique offers a proxy for the more difficult direct measurement of the noise temperature, for example via the measurement induced dephasing of a qubit\cite{Hatridge2013}. Nevertheless, we can infer the directional amplifier noise performance by comparing to the single gain process of the JPC, which has been previously shown to be nearly quantum-limited\cite{Hatridge2013}.  As shown in Fig.~4C, we observe NVR only at the outputs which have gain (I and V) and not at the isolated input port (S). The measured NVR for directional gain agrees to within $1$~ dB with the associated single pairwise coupling with the same gain, although there are slight shifts of the center frequencies. This indicates that the noise performance of the directional amplifier is essentially as quantum-limited as the conventional non-directional phase-preserving amplifier mode of the JPC.

Finally, we examine the behavior of the device as a function of the conversion coefficient (Fig.~5). As detailed in the theory section, the conversion process must dominate for the amplifier to be directional. The dependence of two representative scattering parameters $s_{\mb\mb}$ and $s_{\ma\mb}$ corresponding to input match and directional gain, respectively, are shown for selected conversion coefficients. As expected, the magnitude of all scattering parameters rises as the conversion coefficient decreases, with complete loss of input match and even reflection gain being observed once $C$ falls below a certain threshold (here $C=0.95$, matching the expected value described in the theory section). This threshold rises with the amplifier gain; we have chosen a directional gain of $14$~dB in order to retain sufficient input match. In general, to achieve a single-stage directional amplifier with high forward gain while retaining a matched input, one requires, surprisingly, a nearly perfect converter as the key element.

\section{Discussion and Conclusions}

We have successfully realized a Josephson circuit that performs the functions of both a circulator and a directional amplifier. Each function is determined by a specific pump configuration consisting of three pump frequencies, amplitudes and phases.  To our knowledge, this work represents the first successful implementation of a Josephson microwave circulator, a device of great practical and theoretical interest. Since the circuit is minimally composed of purely dispersive elements, we expect it to be essentially noise-less, unlike in previous proposals \cite{Kamal2010, Lehur2010, Kerckhoff2015}. This same low-loss construction results in nearly quantum-limited noise performance of the circuit when operated as a directional amplifier. Additionally, the simplicity of the device can be an advantage when compared to previous Josephson junction based directional amplifier implementations such as voltage-biased DC-SQUIDs (Superconducting QUantum Inteference Devices) \cite{Clarke2006}, non-linear superconducting transmission lines \cite{HoEom2012, Bockstiegel2014, OBrien2014,Martinis2015}, and coupled JPCs \cite{Abdo2014}.

Our results are in good qualitative agreement with theory, but some discrepancies remain which we attribute to neglect of higher order terms in the theory, as well as imperfection in our control of the relative phase of the pumps. Further work is needed to analyze the dynamic range characteristics, off-resonant response, and higher order mixing products. Our present implementation can be optimized, for example by eliminating the cascaded microwave hybrids. Their elimination by further microwave engineering can render the device fully planar, and will also remove potential sources of loss and mismatch which degrade the device performance. This methodology, combined with the absence of any large magnetic fields, has excellent prospects for on-chip integration with standard circuit QED \cite{ Clarke2008, Devoret2013, Girvin2015} systems, and in other cryogenic measurements using microwave signals, such as kinetic inductance detectors\cite{Day2003}, dispersive magnetometers\cite{Hatridge2011}, and quantum nano-mechanical resonators\cite{Teufel2011}.

Our work also demonstrates that different kinds of non-reciprocal devices can be fundamentally linked. At the basic level, directionality in our devices arises from the non-reciprocal phase shift that signals acquire via a parametric two-mode interaction. This has important implications for the design of future directional devices. An open question that remains is what are the minimal number and nature of couplings needed to produce a desired non-reciprocal scattering matrix. As an example, still more exotic amplifiers such as a directional phase-sensitive amplifier can also potentially be realized along the same lines as in our experiment by the same three modes coupled via a different set of parametric pumps\cite{Metelmann2015}. Furthermore, these two-body interactions can be switched on the fly with a switching time limited by the bandwidth of the device. This results in a flexible device, which can be used to build more complicated signal routing schemes. As an example, we believe these in-situ switchable directional elements could be the basis of a truly quantum switch matrix/gain medium, as pursued in other quantum information platforms\cite{Hucul2015}.


\begin{acknowledgments}
We are indebted to Leonardo Ranzani for his in-depth presentation at Yale of the graph theory of non-reciprocity. We also acknowledge useful discussions with Jos\'{e} Aumentado, Alexander Blais, Aashish Clerk and Archana Kamal. This work was supported by the Army Research Office grant W911NF-14-1-0011. Support from the Air Force Office of Scientific Research is gratefully acknowledged.  Facilities use was supported by the Yale Institute of Nanoscience and Quantum Engineering under National Science Foundation grant MRS 119826.
\end{acknowledgments}

\clearpage
\begin{figure}
\includegraphics{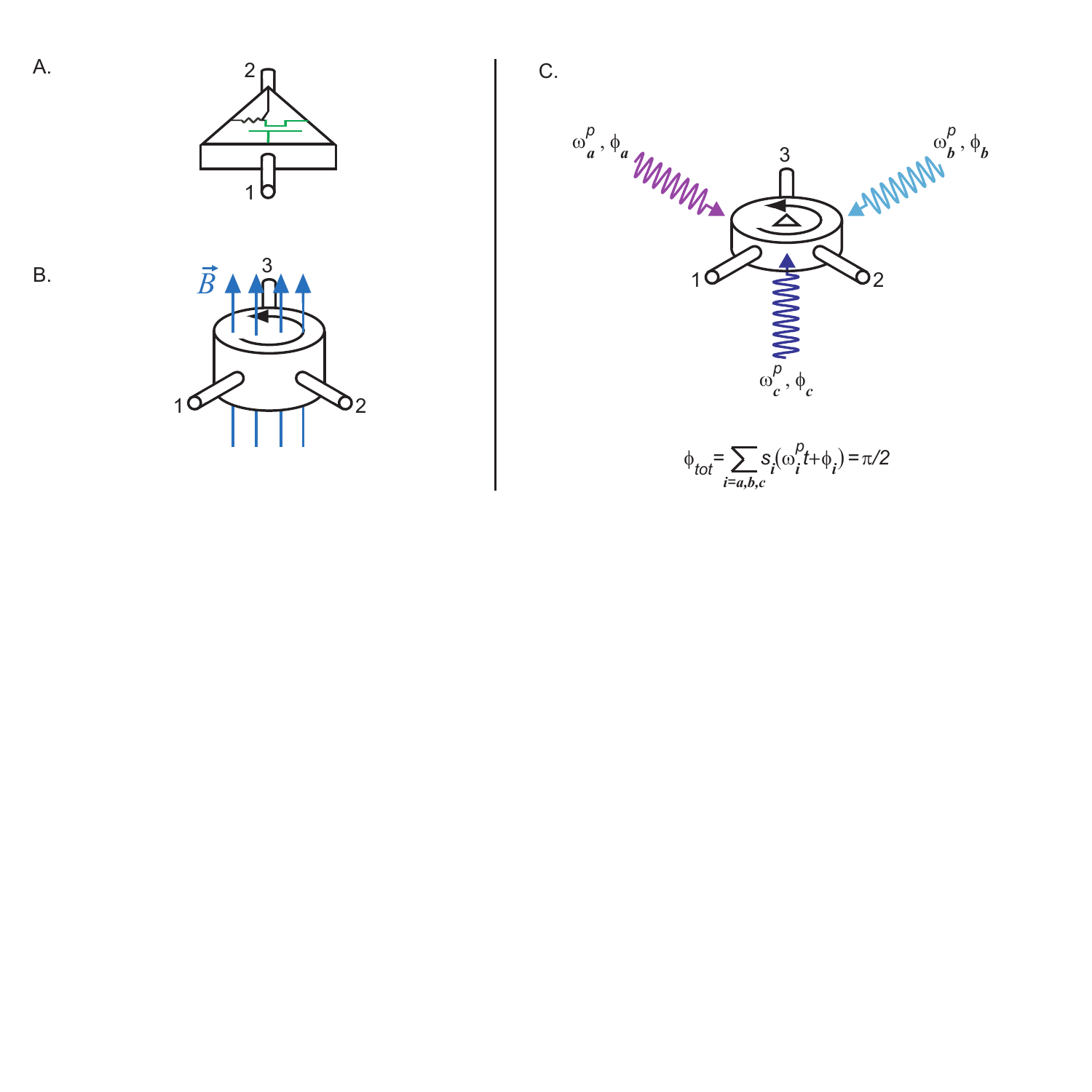}
\end{figure}
\clearpage
\textbf{Figure 1.}
Comparison between two usual microwave directional elements, the directional field-effect-based amplifier (A) and the ferrite circulator (B), and the Josephson triple-pumped reconfigurable circulator/directional amplifier (C) which is the subject of this article. In A, directional amplification with semiconductor devices derives from the field-effect transconductance between the gate (port $1$) and the drain-source (port $2$) circuits, while in B the circulation from port $1$ to port $2$ to port $3$ is obtained from the non-reciprocal property of a ferrite material biased by a magnetic field $\vec{B}$. The different functions of the devices in A and B can be performed by the unique device in C based on a Josephson Ring Modulator coupled to three microwave modes $\ma$, $\mb$, and $\mc$ which form ports $1$, $2$, and $3$. The frequency and phase of three pump tones determines which function is implemented. The directionality of the device is set by the coefficients $s_{i}=\pm1$ entering into the total phase $\phi_{tot}$.

\clearpage
\begin{figure}
\includegraphics{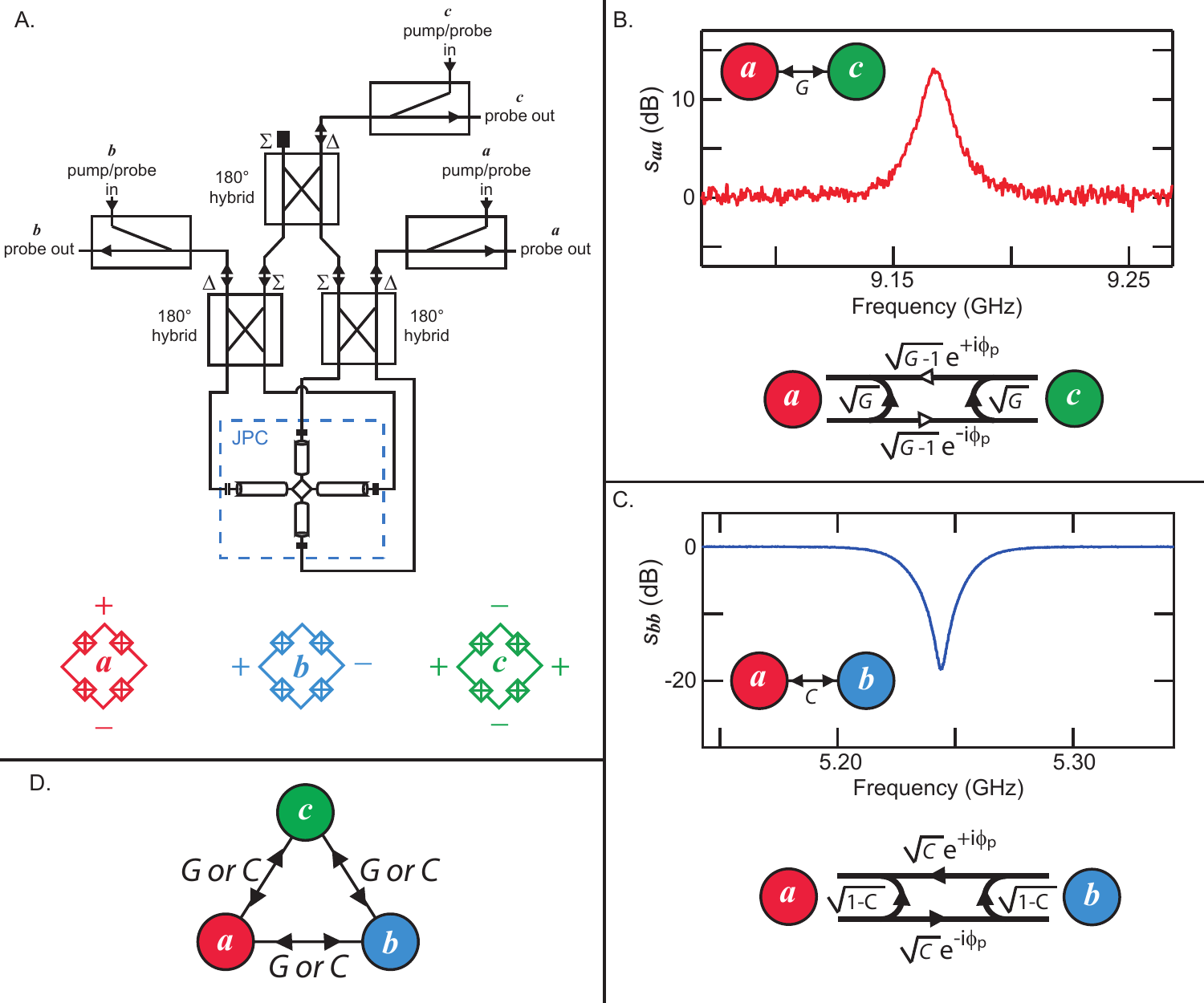}
\end{figure}
\clearpage
\textbf{Figure 2.}
A) Schematic of experimental setup: The Josephson Parametric Converter (JPC) is composed of two $\lambda/2$ microwave resonators with dissimilar resonance frequencies that meet in a central Josephson Ring Modulator consisting of four Josephson junctions forming a loop threaded by a magnetic flux close to a half-flux-quantum. The device has three orthogonal modes with spatial mode patterns depicted below the circuit schematic. Two modes ($\ma$ and $\mb$) are excited by driving each individual resonator through the differential ports of the lower $180^{\circ}$ hybrids. The third mode ($\mc$) is excited by a common drive applied to both resonators via the cascade of three hybrids as shown. Each mode is addressed with a probe tone at the mode frequency, and a pump tone which can couple the other two modes with either a gain or a conversion process. The probe and pump are combined at room temperature, and travel down one line of the dilution refrigerator into the weakly coupled port of a directional coupler at base which serves to separate the input of the JPC from the output.
B) Photon gain amplification process in  which modes $\ma$ and $\mc$ are coupled by a pump applied to mode $\mb$ at frequency $\omega_\ma+\omega_\mc$. The scattering coefficient $s_{\ma\ma}$, plotted versus probe frequency, shows a characteristic lorentzian curve with gain $G=13$~dB. A graph representation of the scattering matrix is also given. The unfilled arrows denote transmission of signals with phase conjugation.
C) Unity photon gain conversion in which modes $\ma$ and $\mb$ are coupled by a pump applied to mode $\mc$ at frequency $\omega_\ma-\omega_\mb$. The graph representation of the scattering matrix is shown, and $s_{\mb\mb}$ is plotted versus probe frequency. $s_{\mb\mb}$ shows a $-18$~dB dip at $\omega_{\mb}$ where photons have been converted from $\omega_\mb$ to $\omega_\ma$ with conversion efficiency $C=0.98$.
D) Diagram of all possible pairwise processes in the JPC: Any two modes of the JPC (represented by colored circles) can be coupled via gain ($G$) or conversion ($C$) by a pump applied to the third mode at the appropriate frequency.

\clearpage
\begin{figure}
\includegraphics{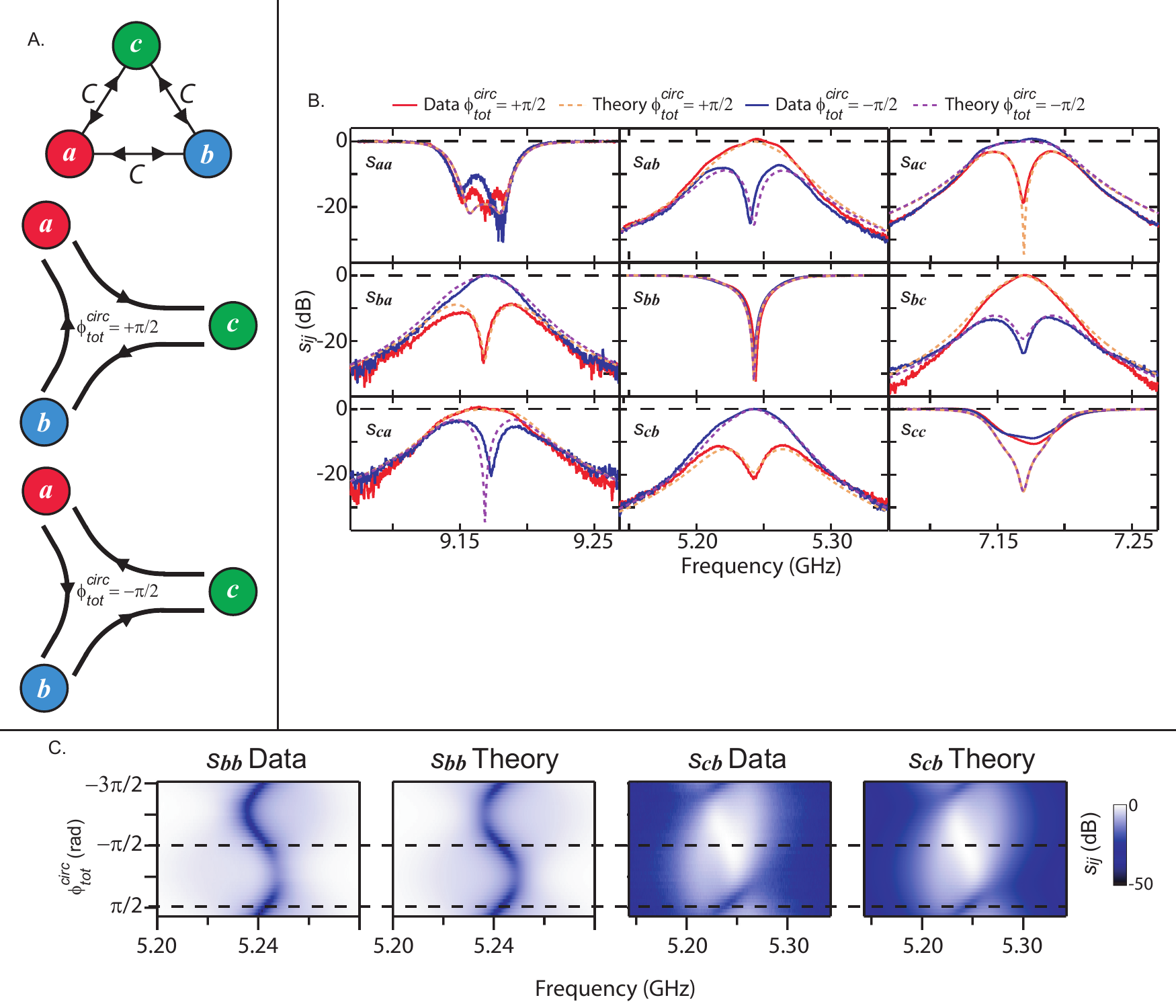}
\end{figure}
\clearpage
\textbf{Figure 3. Circulator}
A) Pump configuration and graphical scattering matrix for a circulator. Linking all pairs of modes via conversion processes realizes a clockwise or counter-clockwise circulator when the total pump phase $\phi_{total}^{circ}= \pi/2$ and $-\pi/2$ respectively.
B) Scattering parameters versus frequency for the clockwise (red) and counter-clockwise (black) circulator, yielding $1$~dB insertion loss and $-10.5$~dB isolation over an $11$~MHz bandwidth.  Theory curves for the clockwise (dashed yellow) and counter-clockwise (dashed violet) circulator are superimposed.
C) Representative scattering parameters (one reflected scattering parameter and one off-diagonal scattering parameter) as a function of phase. Cuts at $\pi/2$ and $-\pi/2$ give the two working points explicitly plotted above, and we see a smooth and symmetric transition between the points. The data and theory are in excellent qualitative agreement.

\clearpage
\begin{figure}
\includegraphics{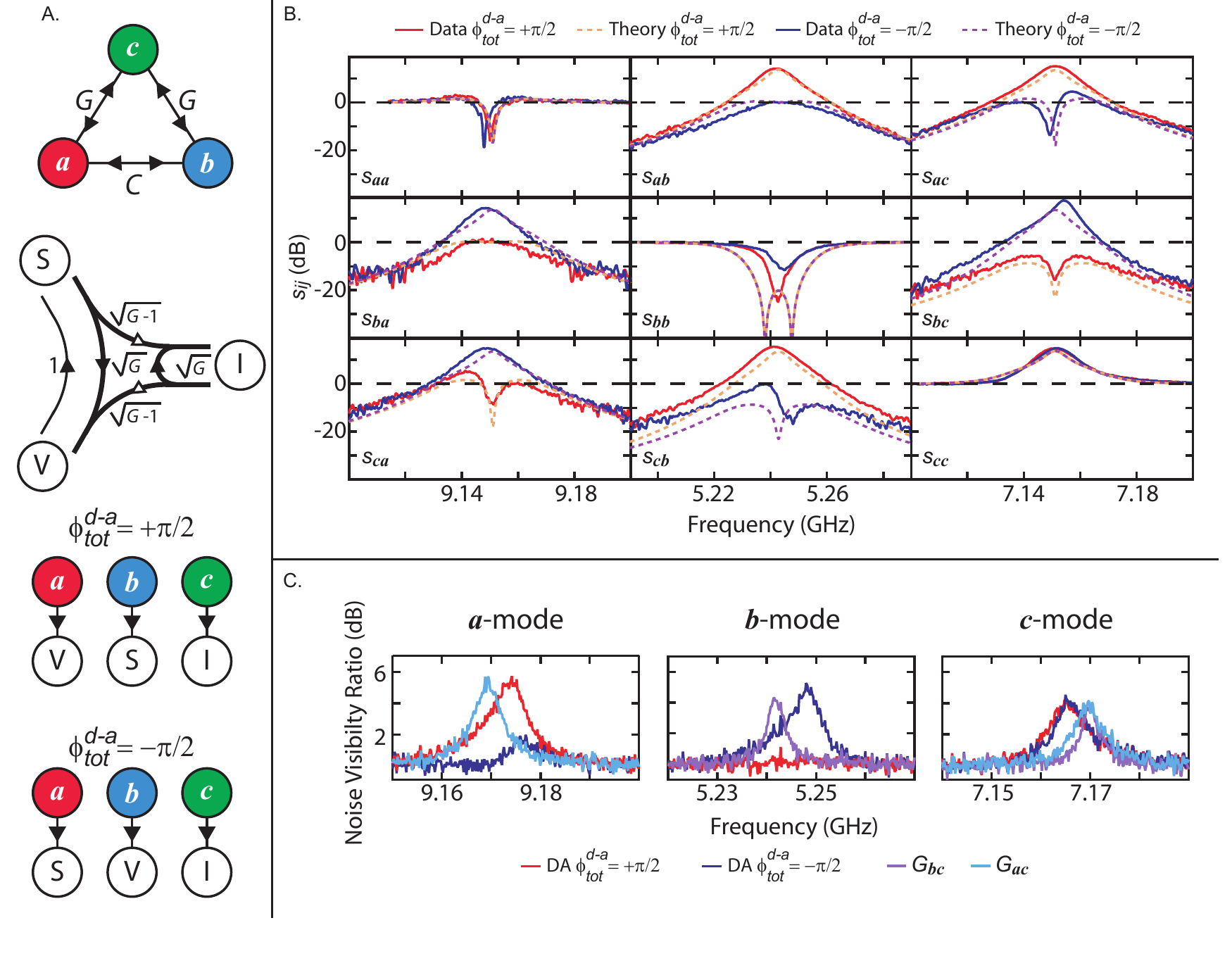}
\end{figure}
\clearpage
\textbf{Figure 4. Directional Amplifier}
A) (Top) Schematic of pump configuration: two pairs of modes are coupled pairwise via gain processes and the third via unity-gain photon conversion. (Middle) Graphical representation of the scattering matrix: the three ports are named for the roles they perform in the amplification process. The signal (S) port serves as the directional amplifier input and is combined via phase-preserving amplification with the idler (I) port, which corresponds to the directional amplifier output.  The vacuum (V) port does not participate in amplification but instead is transmitted with unity gain back to the signal port. For clarity, all zero amplitude scattering parameters are omitted. The unfilled arrows denote transmission of signals with phase conjugation. (Bottom) Map between physical ports and their roles. The pump phase and the choice of which modes are coupled via gain/conversion processes, controls how the physical ports of the JPC are mapped onto the conceptual ports of the directional amplifier. The explicit mapping for the gain and conversion couplings used in the experiment for $\phi_{tot}^{d\text{-}a}= \pi/2$ and $\phi_{tot}^{d\text{-}a}=-\pi/2$ are shown.
B) Scattering parameters versus frequency for both the $\phi_{tot}^{d\text{-}a}= \pi/2$ (red) and $\phi_{tot}^{d\text{-}a}=-\pi/2$ (black) directional amplifier are shown along with superimposed theory (dashed yellow and violet). The amplifier shows $14$~dB of gain and an $11$~MHz bandwidth.
C) Noise visibility ratios for the three ports of the directional amplifier are plotted for both $\phi_{tot}^{d\text{-}a}= \pi/2$ (red) and $\phi_{tot}^{d\text{-}a}=-\pi/2$ (black). They are compared to the noise seen from the individual gain processes between modes $\ma$ and $\mc$ (light blue) and modes $\mb$ and $\mc$ (violet). The noise visibility ratios agree for all ports to within $1$~dB.
\clearpage
\begin{figure}
\includegraphics{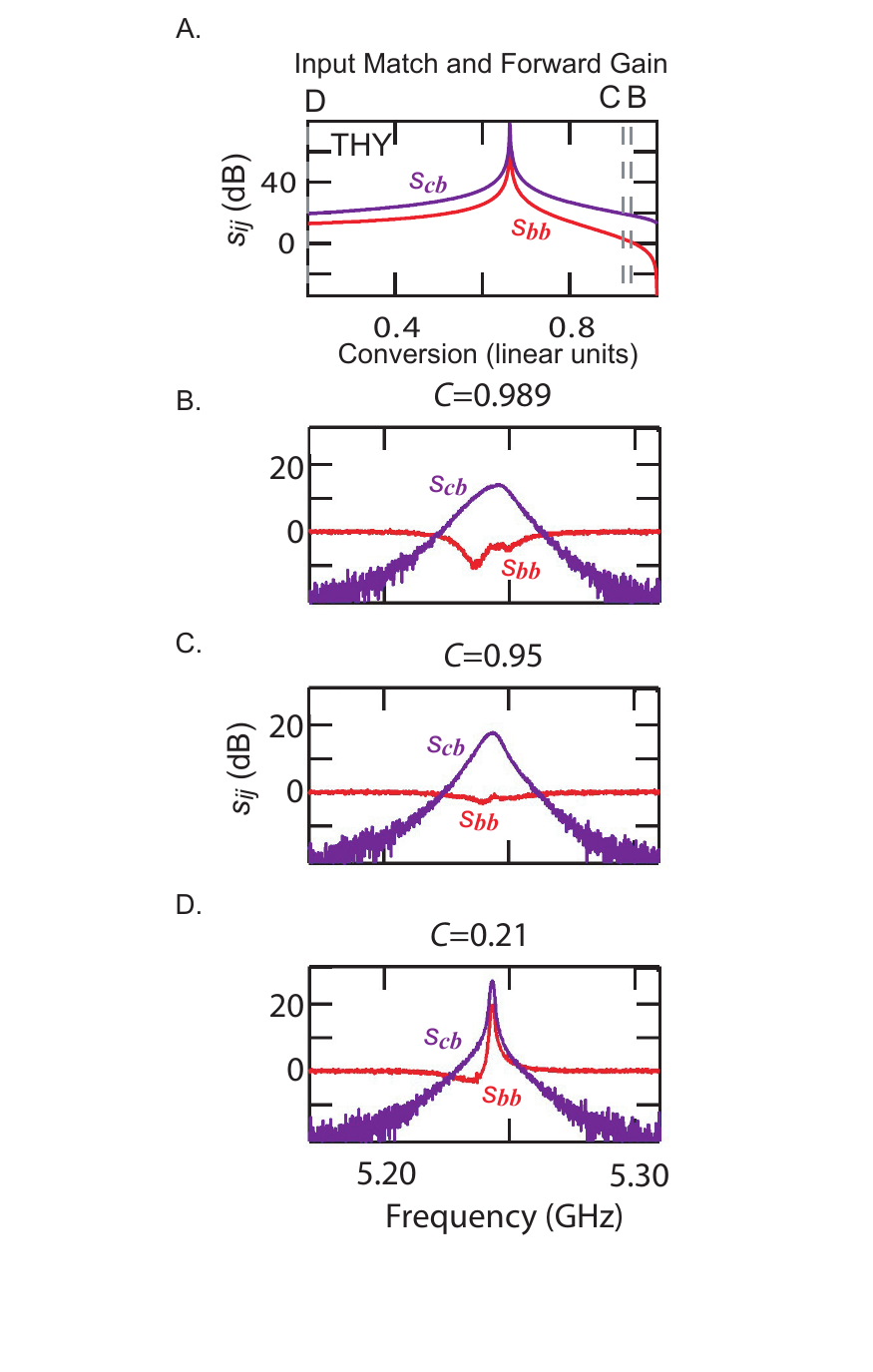}
\end{figure}
\clearpage
\textbf{Figure 5. Dependence of the amplifier directionality on the conversion coefficient.}
A) Calculated scattering parameters for the forward gain $s_{\mc\mb}$ and one of the absorptive input ports $s_{\mb\mb}$ are plotted as a function of conversion with the two single-pump gains $G_{\mb\mc}=12$~dB and $G_{\ma\mc}=13$~dB. For high conversion we see the desired directional gain and input match. As the conversion falls, the match degrades, crossing a threshold at $C=0.95$ past which there is gain on the reflected port. This degradation of directionality (despite the corresponding increase in the forward gain $s_{\mc\mb}$) demonstrates the necessity of high conversion. B, C, D) Measured scattering parameters at $C=0.989$, $0.95$, and $0.21$ showing the expected peaks in both $s_{\mc\mb}$ and $s_{\mb\mb}$ for low conversion transitioning into the desired absorption in $s_{\mb\mb}$ as conversion increases.
\clearpage

\section{Supplementary Material}
\subsection{JPC Fabrication}
The JPC was fabricated on $300$~$\mu$m silicon using the conventional Dolan bridge technique defined using electron beam lithography followed by double-angle aluminum deposition. It consists of two perpendicular $300$~$\mu$m wide $\lambda/2$ microstrip resonators of lengths $4.68$~mm and $9$~mm, each terminated by microstrip gap coupling capacitors ($28$~fF for mode $\ma$ and $32$~fF for mode $\mb$). The coupling capacitors set the energy decay rates $\kappa_\ma$, $\kappa_\mb$, and $\kappa_\mc$ given in the main text. The Josephson Ring Modulator has four $1.7$~$ \mu$A Al/AlOx/Al junctions in a loop shunted by four $3.6$~$ \mu$A Al/AlOx/Al junctions which lift the hysteric behavior of the ring and allow for frequency tuning with an external applied magnetic field.

\subsection{Pairwise gain and conversion characteristics}
Figures S1 and S2 show selected scattering matrix elements for the individual pairwise gain and conversion processes that form the circulator and directional amplifier.  The pumps were tuned individually, with care taken to minimize frequency offsets in the different pairwise processes that share a mode, while maintaining the proper frequency summation condition.

The circulator consists of three pairwise conversion processes (Fig.~S1). We tuned the pump powers to achieve large, well matched conversion coefficients for each pairwise process. Then we simultaneously turned on all three processes and fine-tuned the individual pump powers to maximizing the input match and reverse isolation of the full circulator. We found best performance with pairwise conversion coefficients of $C_{\ma\mb}=0.97$, $C_{\mb\mc}=0.98$, and $C_{\ma\mc}=0.99$.

The directional amplifier consists of one pairwise conversion process and two pairwise gain processes (Fig.~S2). A premium is placed on maximizing the conversion coefficient which limits the achievable directional gain of the device as discussed in the main text. Therefore we first optimized the conversion process and found the conversion coefficient limited to $C_{\ma\mb}=0.998$. The slight double-dip seen in $S_{\ma\ma}$ illustrates the effect of higher order terms that get excited via our attempts to convert as strongly as possible. The pairwise gain processes were then set to $G_{\ma\mc}=12$~dB and $G_{\mb\mc}=13$~dB, chosen to not overwhelm $C_{\ma\mb}$.

\subsection{Directionality as a function of individual gain and conversion processes in the directional amplifier }
Following the methodology in Ref.~\onlinecite{Ranzani2015} we can directly calculate the dependence of the scattering parameters on the individual gain and conversion processes.
Setting the parametric process between modes $\ma$ and $\mb$ to be a conversion process, and the parametric process between both mode pairs $\mb$ and $\mc$, and modes $\ma$ and $\mc$ to be gain processes, and setting $\phi_{tot}^{d\text{-}a}=\pi/2$ we calculate the scattering parameter corresponding to the input match on resonance $S_{\mb\mb}$ in terms of the single pump coupling strengths on $g_{\ma\mb}$,  $g_{\mb\mc}$, and  $g_{\ma\mc}$ and find

\begin{equation}
S_{\mb\mb}=-1+\frac{2(\frac{g_{\ma\mc}^2}{\kappa_{c}\kappa_{a}}-1)}{\frac{g_{\ma\mc}^2}{\kappa_\mc\kappa_\ma}+\frac{g_{\mb\mc}^2}{\kappa_\mb\kappa_\mc}-\frac{g_{\ma\mb}^2}{\kappa_\ma\kappa_\mb}-1}
\end{equation}

The threshold for the loss of directionality corresponds to $S_{\mb\mb}=1$ resulting in the relation

\begin{equation} 
\frac{|g_{\ma\mb}|}{\sqrt{\kappa_\ma\kappa_\mb}}=\frac{|g_{\mb\mc}|}{\sqrt{\kappa_\mb\kappa_\mc}}.
\end{equation}

The single pump coupling strengths can be expressed in terms of the gain and conversion coefficients using $\sqrt{G_{\mb\mc}}=(1+|g_{\mb\mc}|^2/\kappa_{\mb}\kappa_{\mc})/(1-|g_{\mb\mc}|^2/\kappa_{\mb}\kappa_{\mc})$ and $\sqrt{C}=(2|g_{\ma\mb}|/\sqrt{\kappa_{\ma}\kappa_{\mb}})/(1+|g_{\ma\mb}|^2/\kappa_{\ma}\kappa_{\mb})$. Thus we find the threshold for the loss of directionality occurs when $(1-C)<1/G_{\mb\mc}$.

\subsection{Detailed experimental configuration}
Figure S3 shows the circuit diagram of the experimental implementation, including the RF components and generators used at room temperature, as well as the line attenuations in our dilution refrigerator. The components at base temperature are configured as shown in Fig.~2

\clearpage
\begin{figure}
\includegraphics{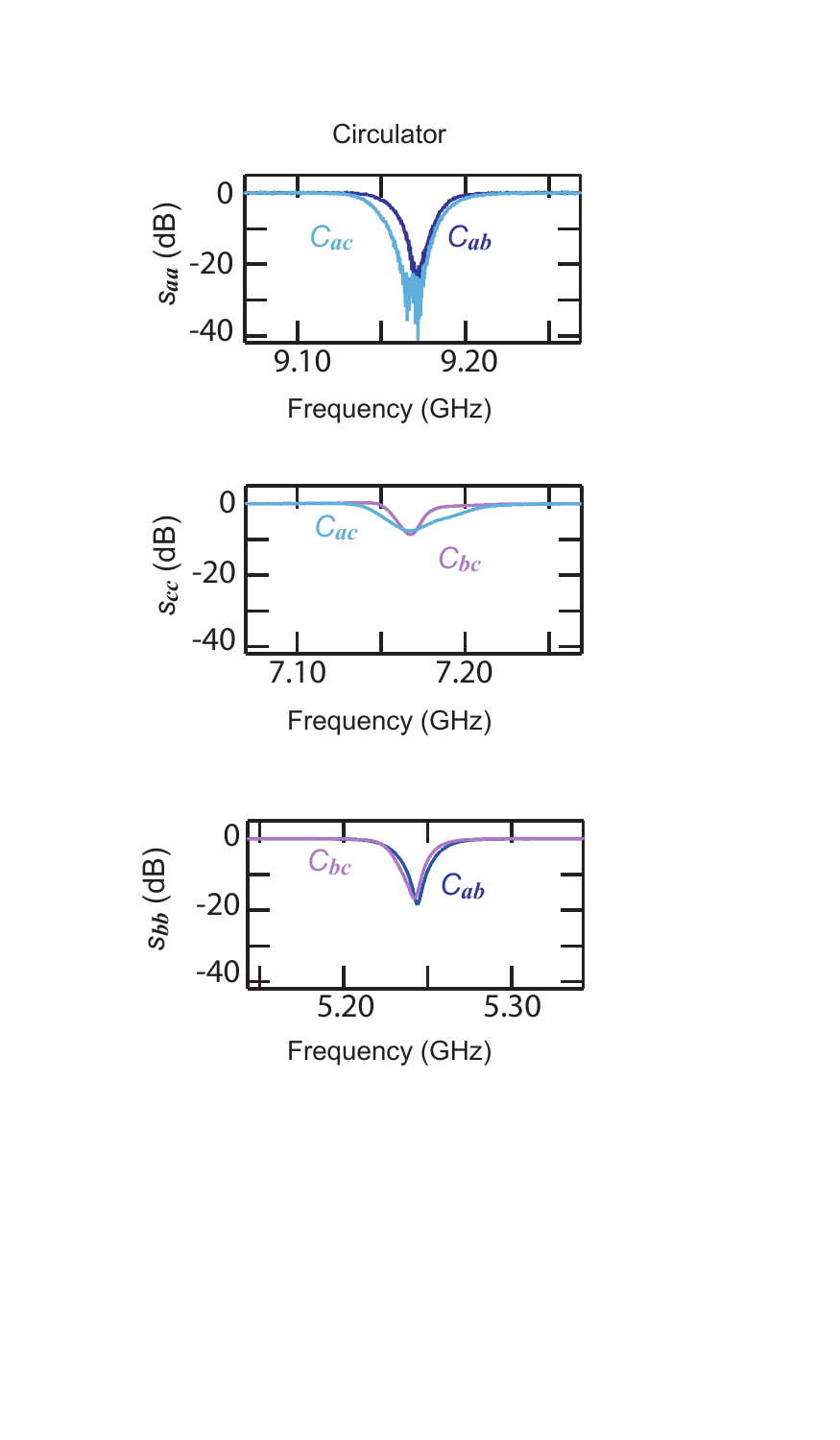}
\end{figure}
\textbf{Figure S1. Pairwise processes in the circulator.}
The measured pairwise conversion curves for all three processes used in the circulator. The symbol $C_{ij}$ indicates a conversion process linking modes $i$ and $j$.  For each port (\ma\, \mb\, \mc), the pump parameters are chosen to match the conversion coefficient and center frequency for both processes involving that mode. We attribute the smaller than expected response of $s_{\mc\mc}$ for both pumps to mismatches in the cascaded hybrid structure.
\clearpage

\begin{figure}
\includegraphics{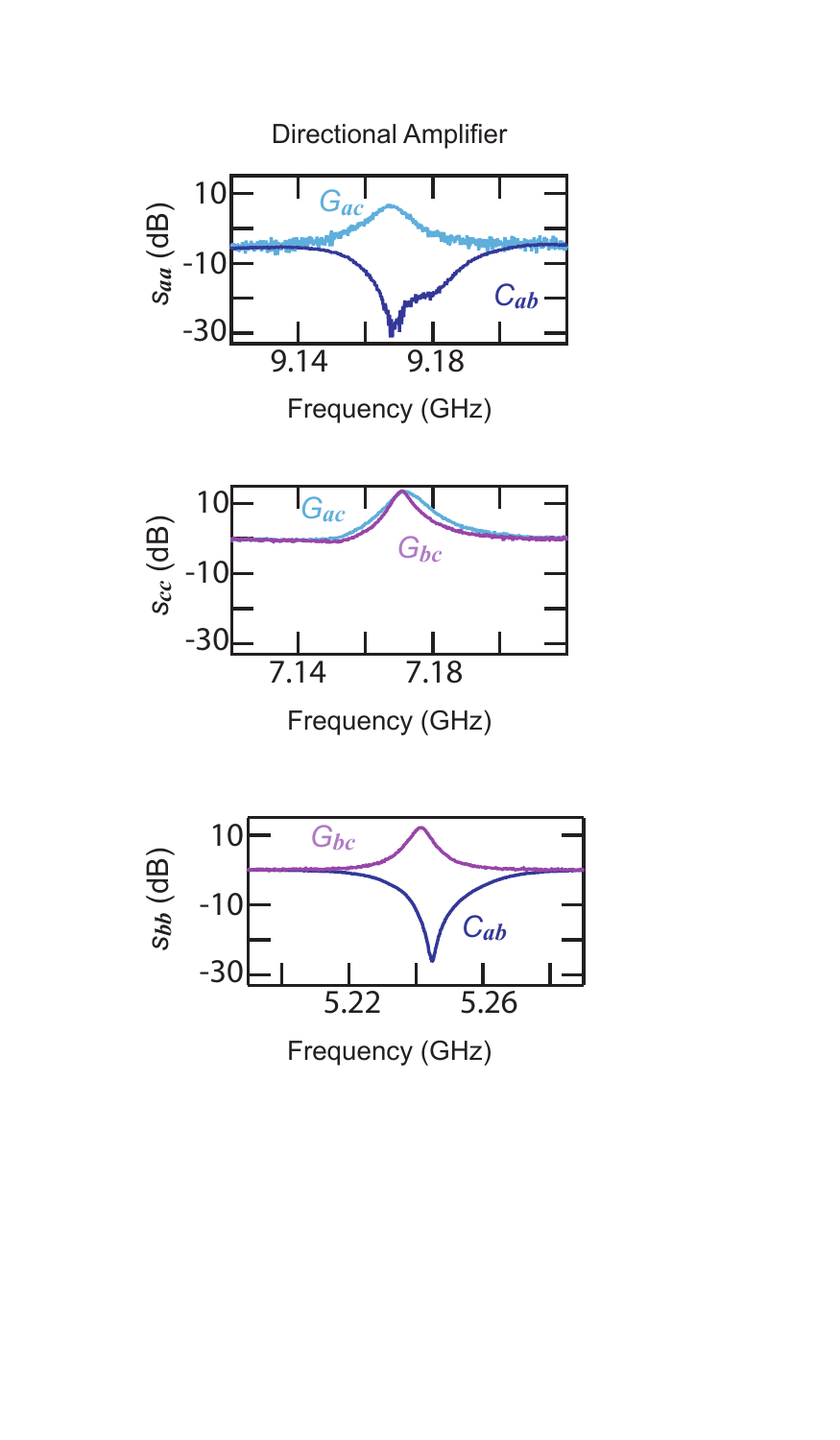}
\end{figure}
\textbf{Figure S2. Pairwise processes in the directional amplifier.}
The measured pairwise conversion and gain curves for all three processes used in the directional amplifier. The symbol $G_{ij}$ indicates a gain process linking modes $i$ and $j$.  For each port (\ma\, \mb\, \mc), the pump parameters are chosen to match the center frequencies for both processes involving that mode.  Additionally, the two photon gains are chosen to match and the conversion is set as close to unity as possible.
\clearpage

\begin{figure}
\includegraphics{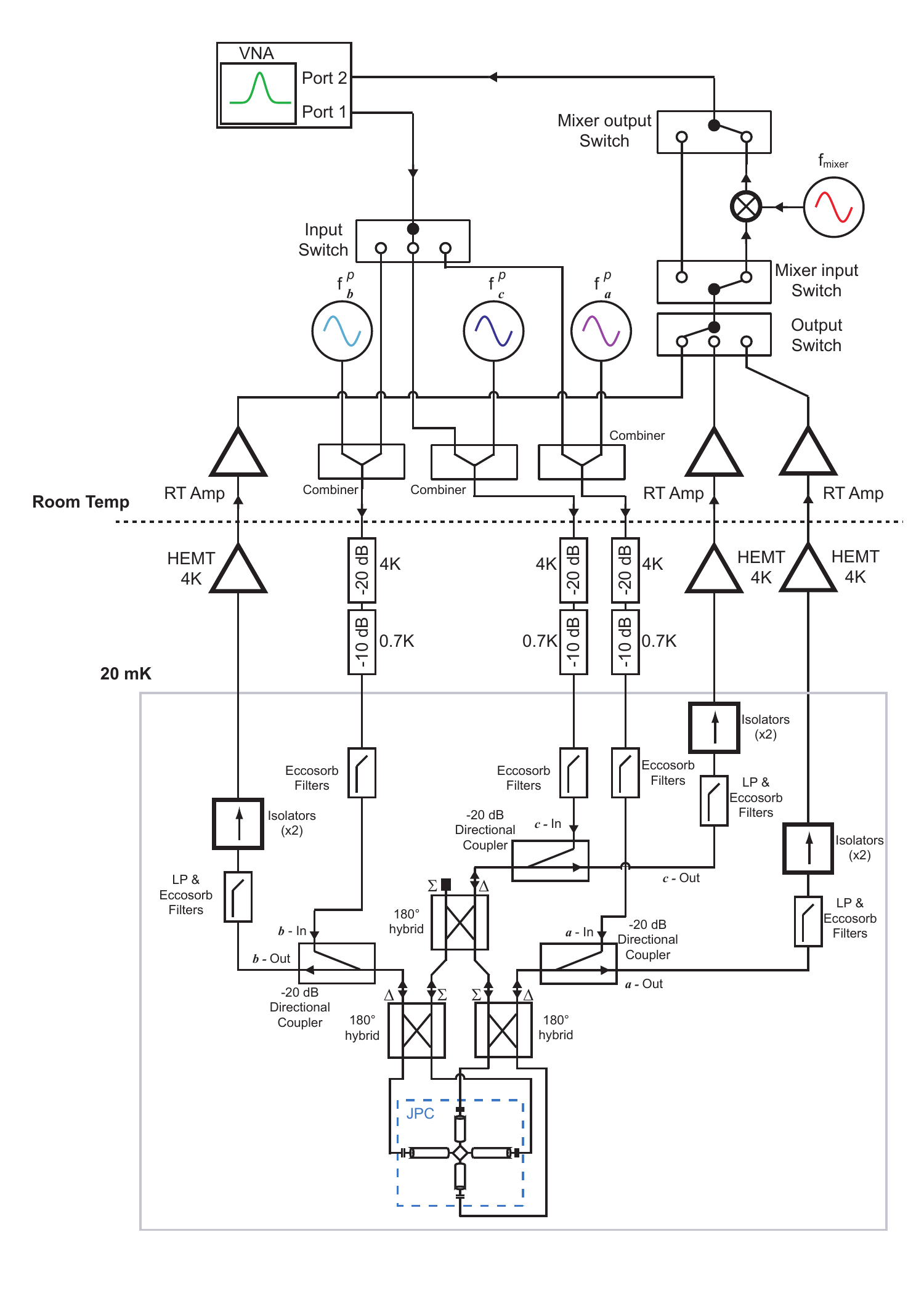}
\end{figure}
\clearpage
\textbf{Figure S3. Detailed circuit diagram of the experimental configuration.}
The Josephson Parametric Converter (JPC) is connected to a cascade of $180^{\circ}$ hybrids, the lower two of which address modes $\ma$ and $\mb$, and the upper which addresses mode $\mc$. Each hybrid is connected to a directional coupler which separates input from output. A directional coupler was chosen for this purpose instead of a circulator due its large operating frequency bandwidth (4-20 GHz). At room temperature, pumps are applied to each input port  via separate generators locked to the same $10$~MHz rubidium atomic clock. The probe tone is sourced from a vector network analyzer (VNA), and a switch is used to pick which input port is addressed. After the switch, the probe is combined with the appropriate pump tone and travels down the fridge through filters and attenuators to the weakly coupled port of the appropriate directional coupler. The output of each port of the JPC travels out of the fridge through a standard set of filters, isolators, and higher stages of amplification. The choice of which output is measured is controlled by a switch at room temperature. If the VNA probe tone and the measured output tone are at different frequencies, the output is directed to a mixer where the output frequency is mixed back to that of the input. If the input and output tones are at the same frequency, the mixer is bypassed. We note all the switches used are specially chosen to terminate the unconnected ports to $50$~$\Omega$.

\clearpage

\end{document}